\documentclass[amsfonts,amssymb,amsmath,aps,prl,notitlepage]{revtex4-1}
\usepackage[colorlinks=true]{hyperref}

\begin{document}

\title{Non-Fierz--Pauli bimetric theory from quadratic curvature gravity on Einstein manifolds}
\author{Yuki Niiyama}
\email{niiyama(at)tap.st.hirosaki-u.ac.jp}
\author{Yuya Nakamura}
\author{Ryosuke Zaimokuya}
\author{Yu Furuya}
\author{Yuuiti Sendouda}
\affiliation{Graduate School of Science and Technology, Hirosaki University, Hirosaki, Aomori 036-8561, Japan}
\date{\today}

\begin{abstract}
We show that, in four-dimensional spacetimes with an arbitrary Einstein metric, with and without a cosmological constant, perturbative dynamical degrees of freedom in generic quadratic-curvature gravity can be decoupled into massless and massive parts.
The massive part has the structure identical to, modulo the over-all sign, the non-Fierz--Pauli-type massive gravity, and a further decomposition into the spin-$ 2 $ and spin-$ 0 $ sectors can be done.
The equivalence at the level of equations of motion allows us to translate various observational bounds on the mass of extra fields into constraints on the coupling constants in quadratic curvature gravity.
We find that the Weyl-squared term is confronting two apparently contradicting constraints on massive spin-$ 2 $ fields from the inverse-square law experiments and observations of spinning black holes.
\end{abstract}

\maketitle

Gravity theories whose Lagrangian contains higher-curvature terms have long been a subject of study since Weyl proposed his conformally invariant theory \cite{Weyl:1919fi} as an alternative to Einstein's General Relativity (GR).
Although there has not been any evidence of GR's failure in passing observational tests, extending gravity theory in a direction to include higher-curvature terms has various theoretical motivations from regularization of energy-momentum tensors on curved spaces \cite{Utiyama:1962sn}, quantum gravity \cite{tHooft:1974toh,Stelle:1976gc}, and the low-energy effective theories of superstrings \cite{Gross:1986iv}.
In this context, the minimally extended class of gravity theories on top of GR, or lowest-order expansion of general $ f(\text{Riemann}) $ gravity, can be described by the action
\begin{equation}
S[g]
= \frac{1}{2\kappa}\,\int d^4 x\,\sqrt{-g}
  \left(R - 2 \Lambda - \alpha\,C_{\mu\nu\rho\sigma}\,C^{\mu\nu\rho\sigma} + \beta\,R^2\right)\,,
\label{eq:action1}
\end{equation}
where $ g_{\mu\nu}$ is the metric tensor, $ R $ the Ricci scalar, $ C_{\mu\nu\rho\sigma} $ the Weyl tensor, $ \Lambda $ a cosmological constant, $ \kappa $ the constant of gravitation, and $ \alpha $ and $ \beta $ the coupling constants characterizing the theory with the dimension of length squared.
In four dimensions, this is the most general parity-preserving action written in terms of the curvature tensors up to quadratic order, since the other possible combinations $ R_{\mu\nu}\,R^{\mu\nu} $ and $ R_{\mu\nu\rho\sigma}\,R^{\mu\nu\rho\sigma} $ can be absorbed into $ R^2 $ and $ C_{\mu\nu\rho\sigma}\,C^{\mu\nu\rho\sigma} $ thanks to the topological nature of the Gauss--Bonnet combination $ \mathcal R_\mathrm{GB}^2 \equiv R_{\mu\nu\rho\sigma}\,R^{\mu\nu\rho\sigma} - 4 R_{\mu\nu}\,R^{\mu\nu} + R^2 $\,.

Since the action \eqref{eq:action1} of the generic quadratic-curvature gravity (QCG) contains terms non-linear in second-order derivatives of the metric, its equation of motion (eom) becomes a fourth-order differential equation, signaling the emergence of extra degrees of freedom (dofs) other than the gravitons in GR.
Indeed, general hamiltonian analyses \cite{Deruelle:2009zk,Kluson:2013hza} imply that the theory contains scalar- and tensor-type (in the three-dimensional sense) extra dofs (see also \cite{Hindawi:1995an}).

When it comes to examining viability of extended theories, the (non-)existence of \emph{ghosts}, dofs with kinetic energy unbounded below, have been playing decisive roles.
In a system where an ordinary dof and a ghost dof coexist, unbounded energy flows from the ghost to the non-ghost dof would occur once the two dofs come into contact with each other, leading to a runaway catastrophe of the system.
Thus the absence of ghost has been usually taken as a criterion for deciding that a given theory is healthy (see, e.g., \cite{Pais:1950za,Woodard:2006nt,Sbisa:2014pzo}).

Meanwhile, one can also argue that the existence of ghosts is not prohibited, at least at the classical level, as long as the ghost decouples from the non-ghost dofs.
As Stelle showed \cite{Stelle:1977ry}, this is what happens when the QCG action \eqref{eq:action1} is expanded up to the second order in metric perturbations around the Minkowski background ($ \Lambda = 0 $):
The linear theory consists of a massless spin-$ 2 $, a massive spin-$ 0 $, and a massive spin-$ 2 $ dofs, the last one being identified to be ghost, but the occurrence of ghost instabilities is evaded since there are no interactions between the dofs.
Following Stelle, various authors have shed light on the decoupling property of massive dofs in general or subclasses of QCG on backgrounds with maximal symmetry \cite{Clunan:2009er,Lu:2011zk,Hyun:2011ej,Tekin:2016vli}, spherical symmetry \cite{Whitt:1985ki,Myung:2013bow}, or Ricci flatness \cite{Lu:2017kzi}.

Decoupling of the dofs does not immediately mean the linear stability of a given system.
Other criteria for the stability, such as avoidance of tachyons, should be met.
If one wants to study linear stability in realistic setups such as asymptotically flat black holes or inflationary cosmology, one has to invoke a dedicated analysis in each setup.
A pioneering work on the stability of Schwarzschild black hole in QCG was done by Whitt \cite{Whitt:1985ki}, and more recently in \cite{Myung:2013doa} and \cite{Lu:2017kzi}.
Possible signatures of the ghost in the primordial gravitational waves were investigated in \cite{Clunan:2009er,Deruelle:2012xv}.

Of course, the situation would change when coupling to matter fields and/or non-linear interactions are taken into account, but, even if there are relevant couplings, it is a non-trivial task to prove the time scale of instability is short enough to threaten the observational consistency of the theory.
Though more severe problems may show up when moving to quantum theories, such as vacuum decay \cite{Sbisa:2014pzo} or pathology of negative probability \cite{Woodard:2006nt}, there is no evidence or urgent necessity for gravity to be quantized.
Thus, in order to draw a robust conclusion, a firm way to go seems to carry out thorough investigations at the linear level in classical theory.

The primary aim of this Letter is to establish an efficient method for analyzing linear QCG in various phenomenologically important spacetimes by unifying the above mentioned results on the decoupling of the dofs and extending them to arbitrary Einstein manifolds.
For this purpose, it is convenient to introduce a Lovelock tensor \cite{Lovelock:1971yv}
\begin{equation}
\mathcal G_{\mu\nu}
\equiv
  R_{\mu\nu} - \frac{R}{2}\,g_{\mu\nu} + \Lambda\,g_{\mu\nu}\,,
\end{equation}
where $ R_{\mu\nu} $ is the Ricci tensor and $ \Lambda $ the cosmological constant appearing in the action \eqref{eq:action1}.
Using $ \mathcal G_{\mu\nu} $\,, \eqref{eq:action1} is rewritten as
\begin{equation}
S[g]
= \frac{\Delta}{2\kappa}\,\int d^4x\,\sqrt{-g}\,
  \left(
   2 \Lambda
   - \mathcal G
   - \frac{\bar\alpha}{2}\,\mathcal G_{\mu\nu}\,\mathcal G^{\mu\nu}
   + \frac{\bar\beta}{2}\,\mathcal G^2
  \right)
  - \frac{\alpha}{2\kappa}\,\int d^4x\,\sqrt{-g}\,\mathcal R_\mathrm{GB}^2\,, 
\label{eq:action2}
\end{equation}
where $ \mathcal G \equiv g^{\mu\nu}\,\mathcal G_{\mu\nu} $\,, $ \Delta \equiv 1 + (8 \beta + 4\alpha/3)\,\Lambda $\,, $ \bar\alpha \equiv 4\alpha/\Delta $\,, and $ \bar\beta \equiv (2\beta + 4\alpha/3)/\Delta $\,.
Then, let us introduce a class of metrics $ \bar g_{\mu\nu} $ which give $ \mathcal G_{\mu\nu}[\bar g] = 0 $.
As announced, the vanishing of $ \mathcal G_{\mu\nu} $ imposes the metric to be Einstein, namely, $ R_{\mu\nu\rho\sigma}[\bar g] = \frac{\Lambda}{3}\,(\bar g_{\mu\rho}\,\bar g_{\nu\sigma} - \bar g_{\mu\sigma}\,\bar g_{\nu\rho}) + C_{\mu\nu\rho\sigma}[g] $\,, and hence $ R_{\mu\nu}[\bar g] = \Lambda\,\bar g_{\mu\nu} $\,.
These Einstein metrics satisfy the vacuum gravitational field equation derived from \eqref{eq:action1}, so we use it as an unperturbed background solution.
In the following, barred quantities such as $ \bar R_{\mu\nu\rho\sigma} $ are to be understood as evaluated on the Einstein background.
The indices of tensors will be raised and lowered using the background metric $ \bar g $\,.

Let us consider metric perturbations $ h_{\mu\nu} \equiv g_{\mu\nu} - \bar g_{\mu\nu} $\,.
We denote the linear perturbation of the Lovelock tensor $ \mathcal G_{\mu\nu} $ as $ E_{\mu\nu}[h] $\,, i.e.,
\begin{equation}
E_{\mu\nu}[h]
= -\frac{1}{2}\,\bar\Box h_{\mu\nu}
  - \bar R_{\mu\rho\nu\sigma}\,h^{\rho\sigma}
  + \bar\nabla_{(\mu} \bar\nabla^\rho h_{\nu)\rho}
  - \frac{1}{2}\,\bar\nabla_\mu \bar\nabla_\nu h
  + \frac{1}{2}\,g_{\mu\nu}\,\left(\bar\Box h
  - \bar\nabla_\rho\,\bar\nabla_\sigma h^{\rho\sigma}
  + \Lambda\,h\right)\,,
\end{equation}
where $ \bar\nabla_\mu $ denotes the background covariant derivative, $ \bar\Box = \bar g^{\mu\nu}\,\bar\nabla_\mu \bar\nabla_\nu $\,, and $ h = \bar g^{\mu\nu}\,h_{\mu\nu} $\,.
Taking the perturbation of \eqref{eq:action2} to the second order in $ h_{\mu\nu} $\,, we obtain the action for $ h_{\mu\nu} $
\begin{equation}
{}^{(2)}S[h]
= \frac{\Delta}{4\kappa}\,\int d^4x\,\sqrt{-\bar g}\,
  \left(
   -h^{\mu\nu}\,E_{\mu\nu}[h]
   - \bar\alpha\,E_{\mu\nu}[h]\,E^{\mu\nu}[h]
   + \bar\beta\,E[h]^2
  \right)\,,
\label{eq:action3}
\end{equation}
where $ E[h] = \bar g^{\mu\nu}\,E_{\mu\nu}[h] $\,.

It is worth mentioning here that the value and sign of the effective gravitational constant is controlled by the factor $ \Delta $ that depends on the parameters and the cosmological constant, a property being observed in \cite{Tekin:2016vli}.
The special case with $ \Delta = 0 $, corresponding to a generalization of the critical gravity \cite{Lu:2011zk} to a model including a scalar dof, will not be considered below but will be discussed elsewhere.

The equation of motion for $ h_{\mu\nu} $ is, as it must, a fourth-order differential equation which reads
\begin{equation}
E_{\mu\nu}[\bar\alpha\,E[h] - \bar\beta\,E[h]\,g + h]
= \frac{\kappa}{\Delta}\,T_{\mu\nu}\,,
\label{eq:eomh}
\end{equation}
where $ T_{\mu\nu} $ is an energy-momentum tensor of matter which is conserved, $ \bar\nabla^\mu T_{\mu\nu} = 0 $.
To gain some insight into the space of solutions of the eom \eqref{eq:eomh}, it is useful to consider the vacuum case, $ T_{\mu\nu} = 0 $.
Then one can deduce from the above factorized form that there are in general two independent types of solutions:
One is the ``Einstein'' mode obeying
\begin{equation}
E_{\mu\nu}[h]
= 0
\end{equation}
and the other ``non-Einstein'' mode obeying
\begin{equation}
\bar\alpha\,E_{\mu\nu}[h] - \bar\beta\,E[h]\,g_{\mu\nu} + h_{\mu\nu}
= 0\,.
\end{equation}
The general solution for $ h_{\mu\nu} $ is given as a linear combination of the Einstein and the non-Einstein modes.

Based on the above argument, we employ a trick \textit{a la} Stelle \cite{Stelle:1977ry} introducing two variables
\begin{equation}
\phi_{\mu\nu}
\equiv
  h_{\mu\nu} + \bar\alpha\,E_{\mu\nu}[h] - \bar\beta\,E[h]\,\bar g_{\mu\nu}\,,
\quad
\pi_{\mu\nu}
\equiv
  -\bar\alpha\,E_{\mu\nu}[h]+\bar\beta\,E[h]\,\bar g_{\mu\nu}\,,
\end{equation}
to rewrite \eqref{eq:action3} into an equivalent second-order action for the two, decoupled, tensor fields
\begin{equation}
S[\phi,\pi]
= \frac{\Delta}{4\kappa}\,\int d^4x\,\sqrt{-\bar g}\,
  \left[
   -\phi^{\mu\nu}\,E_{\mu\nu}[\phi]
   + \pi^{\mu\nu}\,E_{\mu\nu}[\pi]
   + \frac{\mu^2}{2}\,\left(\pi_{\mu\nu}\,\pi^{\mu\nu}-(1-\epsilon)\,\pi^2\right)
  \right]\,;
\quad
h_{\mu\nu}
= \phi_{\mu\nu} + \pi_{\mu\nu}
\label{eq:action4}              
\end{equation}
where $ \pi = \bar g^{\mu\nu}\,\pi_{\mu\nu} $\,, $ \epsilon \equiv 1 - \bar\beta/(4\bar\beta-\bar\alpha) = 9\beta/(2\alpha+12\beta) $ and $ \mu^2 \equiv 2/\bar\alpha = \Delta/(2\alpha) $\,.
The equations of motion for $ \phi_{\mu\nu} $ and $ \pi_{\mu\nu} $ with a matter source
\begin{equation}
E_{\mu\nu}[\phi]
= \frac{\kappa}{\Delta}\,T_{\mu\nu}\,,
\quad
E_{\mu\nu}[\pi] + \frac{\mu^2}{2}\,\left(\pi_{\mu\nu}-(1-\epsilon)\,\pi\,\bar g_{\mu\nu}\right)
= -\frac{\kappa}{\Delta}\,T_{\mu\nu}\,,
\end{equation}
are equivalent to the equation of motion for $ h_{\mu\nu} = \phi_{\mu\nu} + \pi_{\mu\nu} $\,, \eqref{eq:eomh}.
The action \eqref{eq:action4} consists of a GR part for $ \phi_{\mu\nu} $ and a massive gravity (MG) part for $ \pi_{\mu\nu} $\,, which does not enjoy the Fierz--Pauli (FP) tuning, $ \epsilon = 0 $ \cite{Fierz:1939ix}, if $ \beta \neq 0 $.
It can be shown that $ \phi_{\mu\nu} $ possesses massless two tensorial dofs as in GR since the Lagrangian for $ \phi_{\mu\nu} $ is invariant under the gauge transformation $ \phi_{\mu\nu} \rightarrow \phi_{\mu\nu} + 2 \bar\nabla_{(\mu} \xi_{\nu)} $\,, whereas counting of dofs in $ \pi_{\mu\nu} $ amounts to that in the non-FP type MG;
Since there is no gauge symmetry for $ \pi_{\mu\nu} $ due to the presence of the mass term, five dofs associated with a massive tensor and one massive scalar should arise in general \cite{Bengtsson:1994vn}.
Indeed, the transverse-traceless (TT) part of $ \pi_{\mu\nu} $\,, denoted as $ \psi_{\mu\nu} $\,, satisfying $ \bar g^{\mu\nu}\,\psi_{\mu\nu} = \bar\nabla^\nu \psi_{\mu\nu} = 0 $, and the trace $ \pi $ satisfy the equations of motion
\begin{align}
&
\bar\square \psi_{\mu\nu} + 2 \bar R_\mu{}^\rho{}_\nu{}^\sigma\,\psi_{\rho\sigma} - \mu^2\,\psi_{\mu\nu}
= \frac{\kappa}{\Delta}\,T^\mathrm{TT}_{\mu\nu}\,, \label{eq:eompsi} \\
&
\bar\square \pi - m^2\,\pi
= \frac{\kappa}{\epsilon\,\Delta}\,T\,,
\end{align}
respectively, where $ T^\mathrm{TT}_{\mu\nu} $ and $ T $ are the TT part and the trace of $ T_{\mu\nu} $\,, respectively, and $ m^2 \equiv 1/(6\beta) $\,.

So far, we have established that the massless spin-$ 2 $, massive spin-$ 2 $, and massive spin-$ 0 $ dofs in the linear theory of general QCG \eqref{eq:action1} are decoupled on arbitrary Einstein manifolds with and without a cosmological constant $ \Lambda $\,, including all the vacuum solutions in GR.
The benefit it gives us is an efficient way for analyzing QCG in variety of phenomenologically interesting backgrounds, as we shall see an enlightening example shortly.
Before applying the formalism to a particular situation, it would be worth commenting on a peculiar property of \eqref{eq:action4}.
Until now, we have postponed identifying the dofs in the theory as ghosts or non-ghosts because the overall and relative signs in the action depend on the parameters and cosmological constant:
The overall sign of the action can change according to $ \Delta $\,, as we remarked earlier, but, in any case, the GR and MG parts necessarily have opposite signs, resulting in the unavoidable emergence of a ghost.
For $ \Delta > 0 $, the massless spin-$ 2 $ has the healthy sign and the massive spin-$ 2 $ falls into ghost, whereas for $ \Delta < 0 $, the ghost nature of the two dofs is interchanged.
This is in some sense reminiscent of the Higuchi bound for the massive helicity-$ 0 $ graviton in de Sitter \cite{Higuchi:1986py}, a property also being inherent in \eqref{eq:action4}.
Since the thorough investigation of the ghost spectra is somewhat extensive and involved, we plan to present it elsewhere.

Instead, we now turn to the phenomenological aspects of QCG with the aid of the established correspondence to MG, which allows us to translate the experimental bounds on extra massive fields into constraints on the parameters in the QCG action \eqref{eq:action1}.
To do so, we look at the behavior of the massive dofs in laboratory and astrophysical environments with a vanishing cosmological constant $ \Lambda $\,.

Normally, the masses of extra gravitational dofs are requested to be heavy enough not to violate the inverse-square law of gravitational interaction, which the massless graviton should be responsible for.
As Stelle showed \cite{Stelle:1977ry}, in a Minkowski background, the newtonian potential of a point mass $ M $\,, $ \phi_\mathrm N = -GM/r $\,, receives corrections from the QCG massive fields in the form of Yukawa potential, $ \Delta\phi_\mathrm N = \phi_\mathrm N\,(-4\mathrm e^{-\mu r}/3 + \mathrm e^{-m r}/3) $\,, where, for $ \Lambda = 0 $, the spin-$ 2 $ mass is $ \mu^2 = 1/(2\alpha) $ and the factor $ 4/3 $ for the spin-$ 2 $ part corresponds to the van Dam--Veltman--Zakharov discontinuity in MG \cite{vanDam:1970vg,Zakharov:1970cc}, see e.g.\ \cite{Hinterbichler:2011tt} for a review.
The table-top constraint on the deviation from the $ 1/r^2 $ force comes from a torsion-pendulum experiment \cite{Kapner:2006si}, where, for the prefactor $ -4/3 $, the Yukawa-interaction range is constrained as $ \lambda \lesssim 5 \times 10^{-5}\,\mathrm m $ at 95\% confidence level.
This is converted into a \emph{lower} bound on the spin-$ 2 $ mass $ \mu \gtrsim 4 \times 10^{-3}\,\mathrm{eV} $, which implies that the coupling constant $ \alpha $ must be $ < (4 \times 10^{-5}\,\mathrm m)^2 $.
Note that there is no way circumventing this by getting the massive spin-$ 2 $ field to be nearly massless instead.
Indeed, an existence of a nearly massless extra spin-$ 2 $ field would result in an $ \mathcal O(1) $ deviation of a post-newtonian parameter $ \gamma $ \cite{Will:2014kxa} from the GR value of $ 1 $ \cite{Hinterbichler:2011tt}, causing an inconsistency with the stringent constraint $ \gamma - 1 = (2.1 \pm 2.3) \times 10^{-5} $ from the Cassini experiment \cite{Bertotti:2003rm}.

Meanwhile, Brito \textit{et al.}~\cite{Brito:2013wya} analyzed a super-radiant instability of a massive spin-$ 2 $ field around a Kerr black hole, finding that, for the stability of spinning massive black holes at the centers of galaxies, the spin-$ 2 $ mass is bounded \emph{from above} as $ \mu \lesssim 5 \times 10^{-23}\,\mathrm{eV} $.
When applied to the massive spin-$ 2 $ dof in QCG, a constraint $ \alpha \gtrsim (3 \times 10^{15}\,\mathrm m)^2 $ is found;
One immediately notices that there is a gross conflict with the one from the torsion-pendulum experiment.
Apparently, the only way avoiding this conflict is to set $ \alpha = 0 $ from the beginning, ruling out the Weyl-squared correction.
Here we stress that this upper bound on $ \mu $ is distinct from usual bounds on the \emph{graviton mass} (see e.g.\ \cite{Tanabashi:2018oca}), in that this equally applies to any spin-$ 2 $ dof in the theory as long as the dof obeys the eom \eqref{eq:eompsi}.
Moreover, the cause of the instability is not the ghost nature of the spin-$ 2 $, but is simply due to its massiveness.
Note also that this constraint on $ \alpha $ in Minkowski exists whatever the value of $ \beta $ is.

A caveat here would be that the consequences of back reaction from the radiation of ghosts are rather unclear, and possibly it may take an opposite effect to the one from radiation of non-ghosts;
Though we expect some observable effects on the spin of the black holes, a further analysis is required.

In this Letter, we showed that the linear theory of quadratic curvature gravity on arbitrary four-dimensional Einstein manifolds is equivalent to a linear bi-metric theory with a massless and a massive, non-Fierz--Pauli, gravitons.
The latter can be further decomposed into a massive spin-$ 2 $ and a spin-$ 0 $ parts.
Either of the two spin-$ 2 $ dofs has necessarily a negative sign, leading to an unavoidable emergence of a ghost.
Our result has a wide scope of application to analyses of QCG on various background geometries.
Already from the investigations in Minkowski and Kerr space-times, we obtained two apparently contradicting bounds on the coupling constant $ \alpha $\,.
Among others, interesting phenomenology is expected in anisotropic Bianchi cosmologies since the Weyl curvature tensor can quantify the anisotropy of space.
More conservatively, there should be situations even in the non-Einstein spaces, such as realistic cosmology, where the decoupling of the dofs effectively occurs in some asymptotic regimes as studied in, e.g., \cite{Deruelle:2010kf}.

\begin{acknowledgments}
The authors thank Hideki Asada and Nathalie Deruelle for stimulating and helpful discussions.
The work of YS was in part supported by JSPS KAKENHI Grant Number 16K17675.
\end{acknowledgments}

\bibliographystyle{apsrev4-1}
\bibliography{dec}

\end{document}